\documentclass[aps,prl]{revtex4}
\usepackage{epsfig}
\usepackage{graphicx}
\input{epsf}

\begin{document}

\title{Quantum optical coherence tomography of a biological sample}

\author{Magued B. Nasr,$^{1}$\footnote{Email address: boshra@bu.edu} Darryl P. Goode,$^1$
Nam Nguyen,$^2$ Guoxin Rong,$^2$ Linglu Yang,$^2$ Bj\"{o}rn M. Reinhard,$^2$
Bahaa E. A. Saleh,$^1$ and Malvin C. Teich$^{1}$}
\address{$^{\textit{1}}$Quantum Imaging Laboratory,
Departments of Electrical $\&$ Computer Engineering and Physics,
Boston University, Boston, MA 02215
\\
$^{\textit{2}}$Department of Chemistry, Boston University, Boston, MA 02215}

\begin{abstract}
Quantum optical coherence tomography (QOCT) makes use of an
entangled-photon light source to carry out dispersion-immune axial
optical sectioning. We present the first experimental QOCT images of
a biological sample: an onion-skin tissue coated with gold
nanoparticles. 3D images are presented in the form of 2D sections of
different orientations.

\end{abstract}

\maketitle

\section{Introduction}

A number of nonclassical (quantum) sources of light have come to the
fore in recent years~\cite{TeichSaleh89}--\cite{MandelWolf}, but few
practical applications have emerged. One such application is quantum
optical coherence tomography (QOCT)~\cite{Abouraddy, NasrOptExp}, a
fourth-order interferometric optical-sectioning scheme that makes
use of frequency-entangled photon pairs generated via spontaneous
optical parametric down-conversion (SPDC)~\cite{MandelWolf, SPDC1,
SPDC2}. A particular merit of QOCT is that it is inherently immune
to group-velocity dispersion (GVD) by virtue of the frequency
entanglement of the photon
pairs~\cite{Kwiat2}--\cite{NasrQOCTExpPRL}. Conventional optical
coherence tomography (OCT), in contrast, is a second-order
interferometric scheme that provides high-resolution axial
sectioning by employing ultra-broadband
light~\cite{youngquist87}--\cite{Brezinski}. Unfortunately, however,
this leads to GVD, which degrades resolution~\cite{SalehTeich2007}.
Here we present the first experimental QOCT images of a biological
sample: an onion-skin tissue coated with gold nanoparticles.
Three-dimensional images are displayed in the form of 2D transverse
sections at different depths and 2D axial sections at different
transverse positions. The results reveal that QOCT can become a
viable biological imaging technique.

\section{Experimental arrangement}

\begin{figure}[htbp]
\begin{center}
   \centerline{\includegraphics[width=12cm]{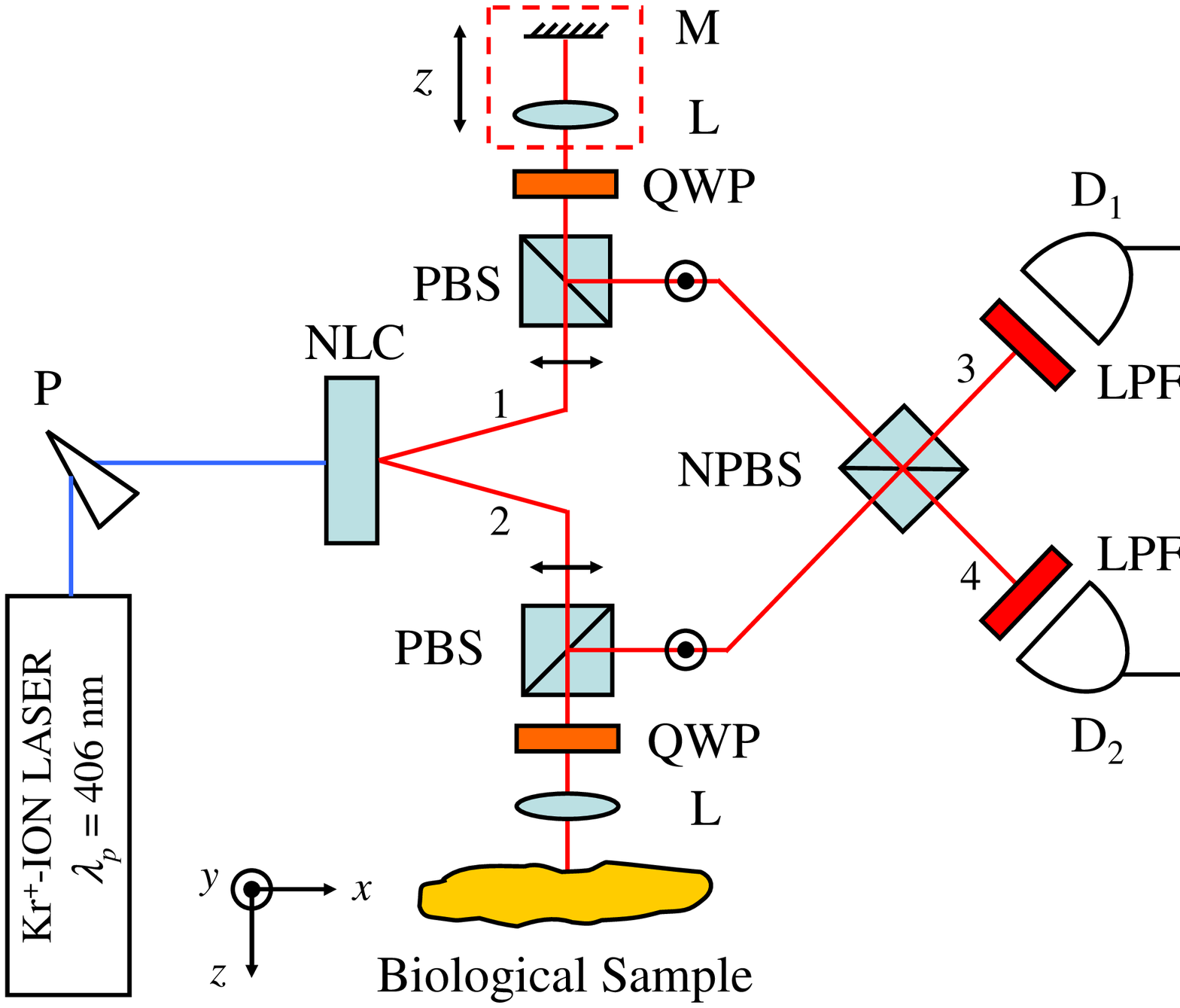}}
    \caption{Experimental arrangement for quantum optical
coherence tomography (QOCT) of a biological sample.}
    \label{ExperimentalQOCT}
 \end{center}
\end{figure}

The details of the QOCT experimental arrangement are provided in
Fig.~\ref{ExperimentalQOCT} (for a detailed review of the theory of
QOCT, the reader is referred to Ref.~\cite{Abouraddy}). A
monochromatic Kr$^{+}$-ion laser operated at a wavelength
$\lambda_{p}$ = 406 nm pumps an 8-mm-thick type-I LiIO$_{3}$
nonlinear crystal (NLC) after passage through a prism (P) and an
aperture (not shown), which remove the spontaneous glow of the laser
tube. A fraction of the pump photons disintegrate into pairs of
down-converted entangled photons
(biphotons)~\cite{MandelWolf, SPDC1, SPDC2}. The entangled photons,
centered about $\lambda =812$ nm, have horizontal polarization with
respect to the optical table and are emitted in a non-collinear
configuration into beams 1 and 2.

The photon in beam 1 is directed to the delay arm, where it is
transmitted through a polarizing beam splitter (PBS) followed by a
broadband quarter-wave plate (QWP), which converts it into a
circularly-polarized photon. It is then focused by an $f=19$-mm
achromatic lens (L), onto a mirror (M). The lens is introduced to
match that in the sample arm (discussed in the next paragraph) to
maintain the indistinguishability of the paths, thereby insuring
interference. The lens-mirror combination is scanned in the
\textit{z} direction during the experiment. The reflected photon
becomes vertically polarized after traversing the QWP for a second
time and it then reflects from the PBS, as illustrated in Fig. 1.

The photon in beam 2 is directed to the sample arm, which contains
an identical set of optical components as the delay arm, with the
exception that the mirror is replaced by a biological sample that
can be scanned in the transverse plane (\textit{x} and \textit{y}
directions). The use of a PBS-QWP pair in the sample arm provides a
factor of four improvement in the number of photons collected
relative to the use of a single non-polarizing beam splitter (NPBS),
as used in earlier demonstrations of QOCT~\cite{NasrOptExp,
NasrQOCTExpPRL}. This factor of four improvement accrues for samples
that do not affect the polarization of the photon. The use of the
lens in the sample arm, in conjunction with a matching lens in the
delay arm, provides a transverse-sectioning capability that was not
previously implemented~\cite{NasrOptExp, NasrQOCTExpPRL}.

The photons returned from the delay arm and the sample arm are each
directed to one of the input ports of a NPBS. Beams 3 and 4, at the
output of the NPBS are directed to single-photon-counting detectors
(EG$\,\&\,$G, SPCM-AQR-15) D$_{1}$ and D$_{2}$, respectively, after
passing through long-pass filters (LPFs) with a cutoff wavelength of
$\lambda$ = 625 nm. A coincidence circuit (denoted $\otimes$)
measures the coincidence rate $C_{x,y}(z)$ between D$_1$ and D$_2$
within a 3.5-nsec time window.

A single QOCT depth scan (A-scan) is obtained by sweeping the
lens-mirror combination in the delay arm in the $z$ direction and
recording the coincidence rate $C_{x,y}(z)$ with the sample fixed at
a particular transverse ($x,y$) position. The collection of all
A-scans for different transverse positions provides the
three-dimensional optical sections for our sample.

If a mirror were to replace the biological sample, the setup
depicted in Fig.~\ref{ExperimentalQOCT} would be equivalent to a
Hong-Ou-Mandel (HOM) interferometer~\cite{HOM} and the coincidence
rate $C_{x,y}(z)$ would trace out a dip whose minimum would occur
when the path lengths of the delay and sample arms were equal. As
such, an A-scan of a biological sample would comprise a collection
of coincidence-rate minima occuring whenever the path length in the
delay arm matches that of a reflecting surface in the sample. These
coincidence-rate minima constitute one class of features that appear
in a QOCT A-scan and comprise the information that is most often
sought in OCT: the depth and reflectance of the internal surfaces
that constitute the sample. Each of these features is associated
with a reflection from a single surface and is immune to the
degradation of axial resolution caused by group-velocity dispersion
(GVD) in the layers above that surface. As will be shown shortly,
these features are used to trace the surface topography of an
onion-skin cell.

A second class of features in the QOCT A-scan arises from cross-interference among the reflection
amplitudes associated with every pair of surfaces and \textit{is}
sensitive to the dispersion characteristics of the media between
them. This class of features generally washes out for photons that are returned from scattering samples~\cite{NasrOptExp}, and we shall not be concerned with it here.

\section{Enhancing the reflectance of a biological sample via coating with gold nanoparticles}

The biological sample investigated in this work was an onion-skin
tissue from a white onion that was coated with spherical gold
nanoparticles to increase its reflectance. Gold nanoparticles have
recently been used as a molecular-specific contrast-enhancement
agent in optical-imaging modalities that rely on the detection of
backscattered light, such as OCT and reflectance-confocal microscopy
(RCM)~\cite{Boppart(2003)}.

We used commercial citrate-stabilized gold nanoparticles with a
nominal diameter of 40 nm, which we modified with two different
surface chemistries. One batch consisted of gold nanoparticles whose
surface was passivated (pegylated) with a monolayer of
self-assembled polyethylene glycols
(O-(2-Carboxyethyl)-O$^\prime$-(2-mercaptoethyl)-heptaethylene
glycols). The second batch of gold nanoparticles contained bovine
serum albumin (BSA) non-specifically attached to the gold surface.
Our gold nanoparticles had a plasmon resonance that peaked at
$\lambda\,$= 527 nm in solution; however, the binding of the gold
particles to the cell surface was accompanied by some agglomeration
of the particles, which red-shifted the plasmon resonance. We
incubated an onion-skin sample into each of the gold-nanoparticle
batches for 48 hours and then rinsed them in distilled water.

\begin{figure}[htbp]
\begin{center}
   \centerline{\includegraphics[width=9cm]{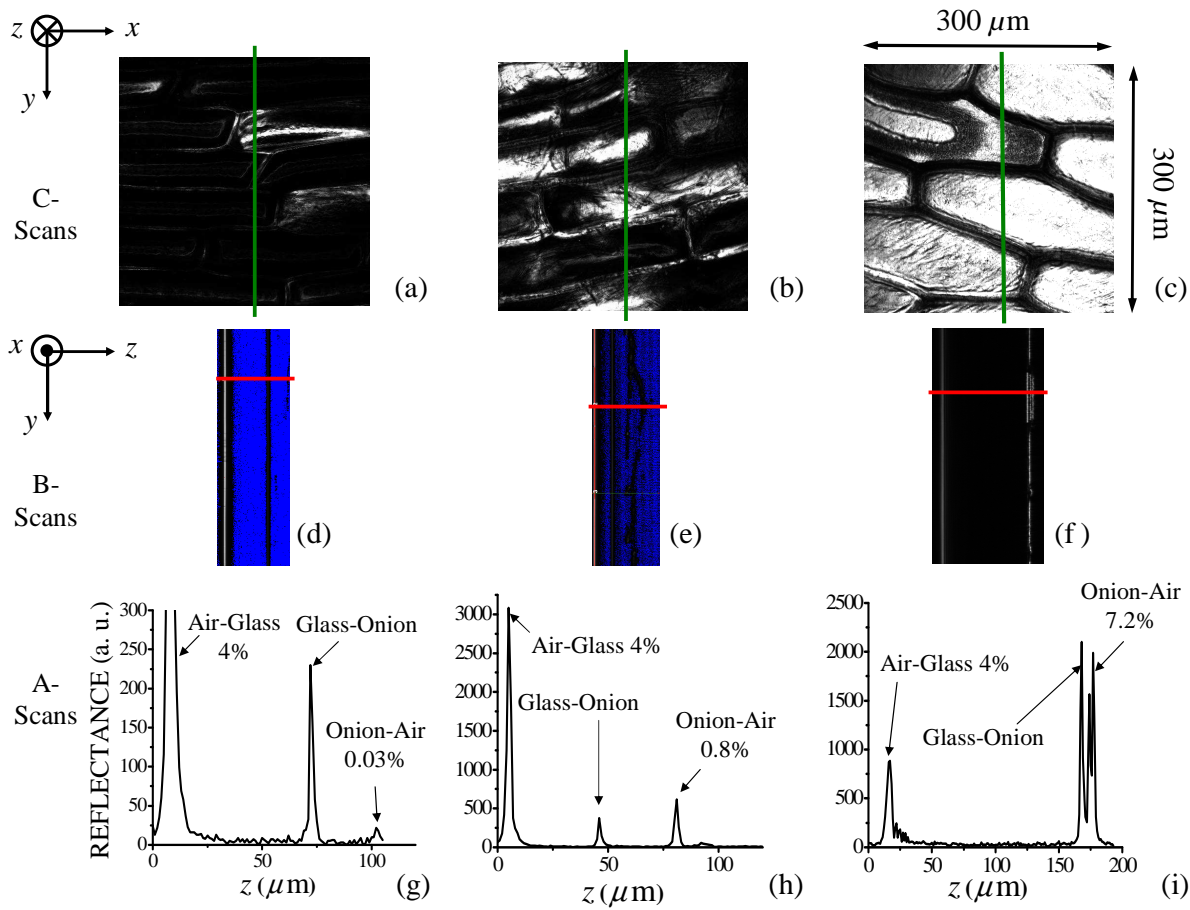}}
    \caption{Reflection confocal microscopy (RCM) scans for different methods of preparing the onion-skin sample.
    Column 1 [(a), (d), (g)]: untreated (bare) onion skin; Column 2 [(b), (e), (h)]: onion skin incubated for
    48 hours in a solution of pegylated gold nanoparticles; Column 3 [(c), (f), (i)]:
    onion skin incubated for 48 hours in a solution of solid gold nanoparticles with bovine serum albumin (BSA) attached to them.}
    \label{RCMOfOnion}
 \end{center}
\end{figure}

We optically investigated the two samples, in addition to an
untreated (bare) sample, using RCM operated at a wavelength of
$\lambda\,$= 632 nm with a 40X objective lens (NA = 0.9). The
results are depicted in Fig.~\ref{RCMOfOnion}. The images in the top
row show 300 $\times$ 300 $\mu\textrm{m}^2$ transverse
(\textit{x\,y}) sections (C-scans) of the onion-skin samples that
are (a) untreated (bare), (b) incubated in the first
gold-nanoparticle batch, and (c) incubated in the second gold
nanoparticle batch. The \textit{x\,y} sections clearly show the
elongated onion-skin cells with dimensions 75 $\times$ 300
$\mu\textrm{m}^2$ in the transverse plane and with a thickness of
about 12 $\mu$m (not shown). Images (d), (e), and (f), appearing in
the middle row, are \textit{y\,z} sections (B-scans) of the same
onion-skin samples that were used to obtain the C-scans, (a), (b),
and (c), respectively. The \textit{y\,z} sections were taken along
the green lines shown in the C-scans in the top row. Finally, in the
bottom row, graphs (g), (h), and (i) display depth scans in the
\textit{z} direction (A-scans) along the red lines shown in the
B-scans, (d), (e), and (f), respectively.

Each of the A-scans exhibits three peaks. The left-most peak arises
from the reflectance at the surface of the microscope cover slip and
corresponds to the 4$\%$ reflectance expected from an air-glass
interface. We used the height of this peak as a reference to
estimate the reflectance of the other peaks. The central peak
corresponds to the glass-onion interface, while the right-most peak
corresponds to the reflection from the surface of the onion-skin
cell. The A-scans in (g) and (h) were obtained by placing the onion
skin on a thin cover slip, whereas a thicker microscope slide was
used to obtain the A-scan in (i), thereby explaining why the
abscissa scale in (i) differs from that in (g) and (h). It is clear
that the onion-skin sample coated with the BSA-functionalized gold
nanoparticles [Fig.~\ref{RCMOfOnion}(c), (f), and (i)] exhibits the
highest reflectance (about 7$\%$). Moreover, the transverse cross
section in image (c) reveals a relatively flatter surface in
comparison with those in images (a) and (b). We believe that this
surface flatness greatly improves the \textit{z} component of the
scattering potential of our sample, which is the measurable quantity
in an OCT or QOCT A-scan~\cite{Zaiat(1995)}.

\section{QOCT optical sectioning of an onion-skin sample}

For the QOCT experiment, we prepared an onion-skin sample coated
with BSA-functionalized nanoparticles, following the method used to
obtain the images in Fig.~\ref{RCMOfOnion}(c), (f), and (i). In this
case, the sample was placed on a 1.25-cm-thick antireflection-coated
glass slab to insure that the back-reflected photons in the sample
arm arose only from the onion skin. We placed the sample in the
setup depicted in Fig.~\ref{ExperimentalQOCT} and obtained a
collection of A-scans at different transverse (\textit{x\,y})
positions of the sample.

At a particular transverse position of the sample, we scanned the
delay-arm lens-mirror combination in the \textit{z} direction over a
range of 30 $\mu\textrm{m}$, using a 1-$\mu\textrm{m}$ step size,
and recorded the coincidence rate $C_{x,y}(z)$ at each step for a
5-second accumulation time. The A-scan obtained thereby was
normalized to the measured value of $C_{x,y}(z)$ at $z = -15\:\mu$m,
which corresponds to the first collected data point, where the path
length of the delay arm did not match the path length to any of the
reflecting surfaces in the sample arm. This choice guaranteed that
each A-scan was not normalized at an interference point.

\begin{figure}[htbp]
\begin{center}
   \centerline{\includegraphics[width=12cm]{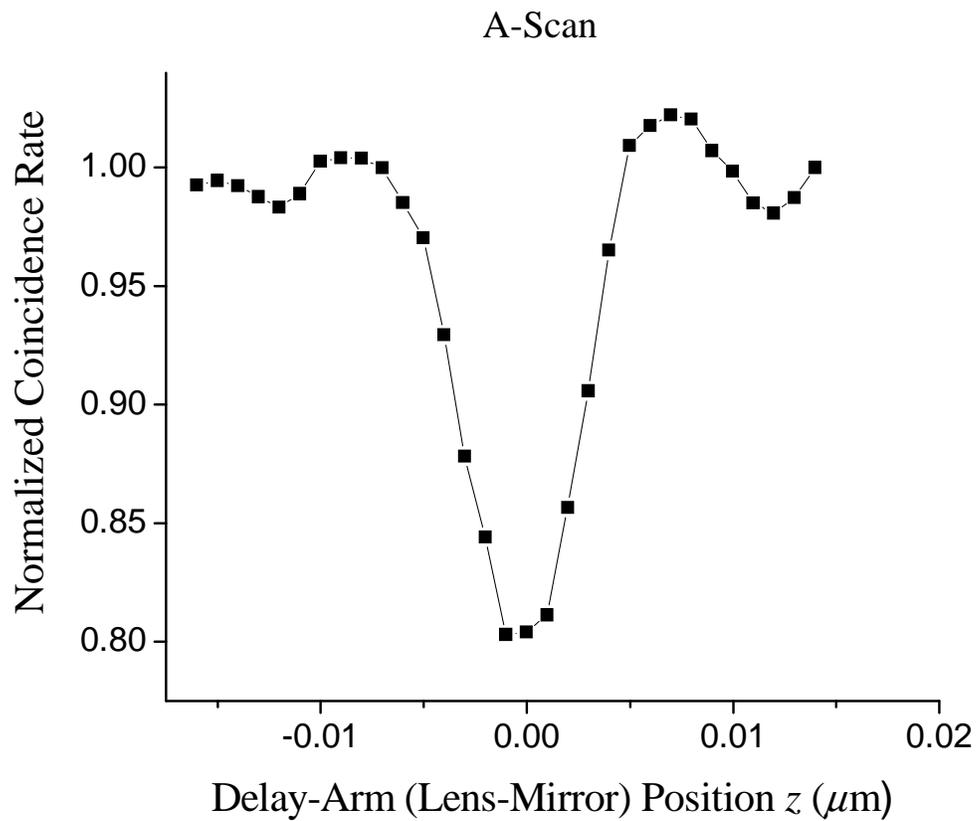}}
    \caption{Typical A-scan showing a coincidence dip that results from reflection from
    the top surface of an onion-skin sample.}
    \label{QOCTofOnion-AScan}
 \end{center}
\end{figure}

The sample was then moved to the next transverse position, with a
step size of 5 $\mu\textrm{m}$ in both the \textit{x} and \textit{y}
directions, and a new normalized A-scan was recorded by repeating
the steps indicated above. A typical A-scan is displayed in
Fig.~\ref{QOCTofOnion-AScan}. The observed dip is a result of
reflection from the top surface of the sample. The width of the dip,
which is $\approx 7.5 \:\mu$m, is in accord with the axial
resolution of the experimental setup, which was measured by using a
mirror in place of the sample. The transverse resolution was
determined by scanning a 5-$\mu$m-diameter pinhole in the focal plane of
the lens; it was estimated to be $\approx 12 \: \mu$m.

The collected A-scans were then used to reconstruct optical sections
of the sample in the form of C-scans and B-scans.

\begin{figure}[htbp]
\begin{center}
   \centerline{\includegraphics[width=9cm]{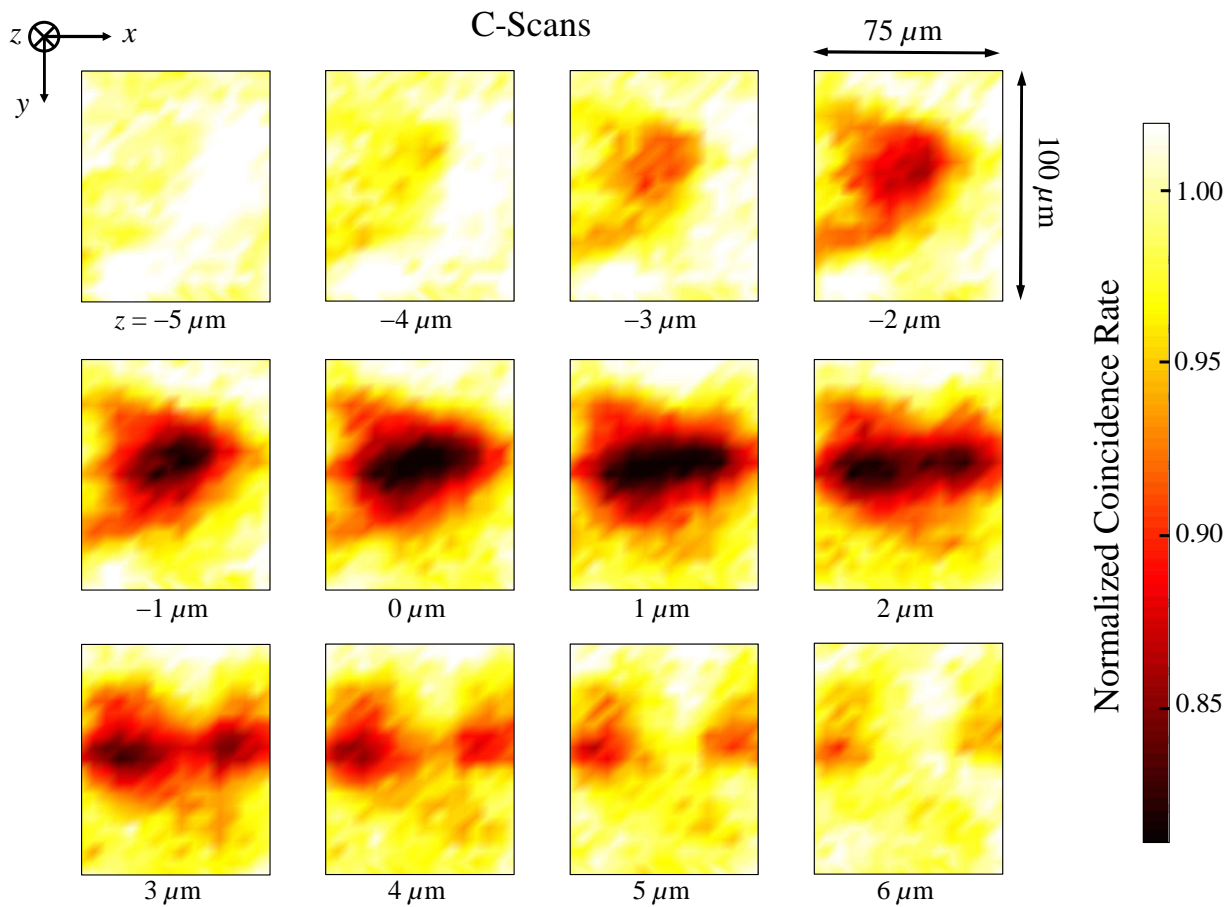}}
    \caption{Two-dimensional transverse (\textit{x\,y}) QOCT sections (C-scans) of an onion-skin sample
    at different axial depths $z$. A strong QOCT signal, representing the detection of a reflecting surface, is indicated by a decrease in the observed coincidence rate.}
    \label{QOCTofOnion-CScan}
 \end{center}
\end{figure}

The C-scans (transverse \textit{x\,y} sections at various depths
$z$) are displayed in Fig.~\ref{QOCTofOnion-CScan}. These
\textit{x\,y} sections are 75 $\times$ 100 $\mu\textrm{m}^2$ each,
and are taken at depth intervals of 1 $\mu\textrm{m}$. The
\textit{z} origin is chosen to approximately coincide with the
surface of the onion-skin cells so that, roughly speaking, negative
\textit{z} positions correspond to sectioning in the air above the
sample, while positive \textit{z} positions correspond to sectioning
within the cells. A strong QOCT signal is indicated by a decrease in
the observed coincidence rate, corresponding to a path-length match
between the delay and the sample arms~\cite{NasrOptExp,
NasrQOCTExpPRL}. The C-scan at $z = 2\:\mu$m clearly shows the
elongated structure that is characteristic of onion-skin cells. The
dimensions of the onion cells observed using QOCT are smaller than
those observed using RCM, probably because of normal sample
variation (smaller-sized onion) or dehydration of the sample.

\begin{figure}[htbp]
\begin{center}
   \centerline{\includegraphics[width=9cm]{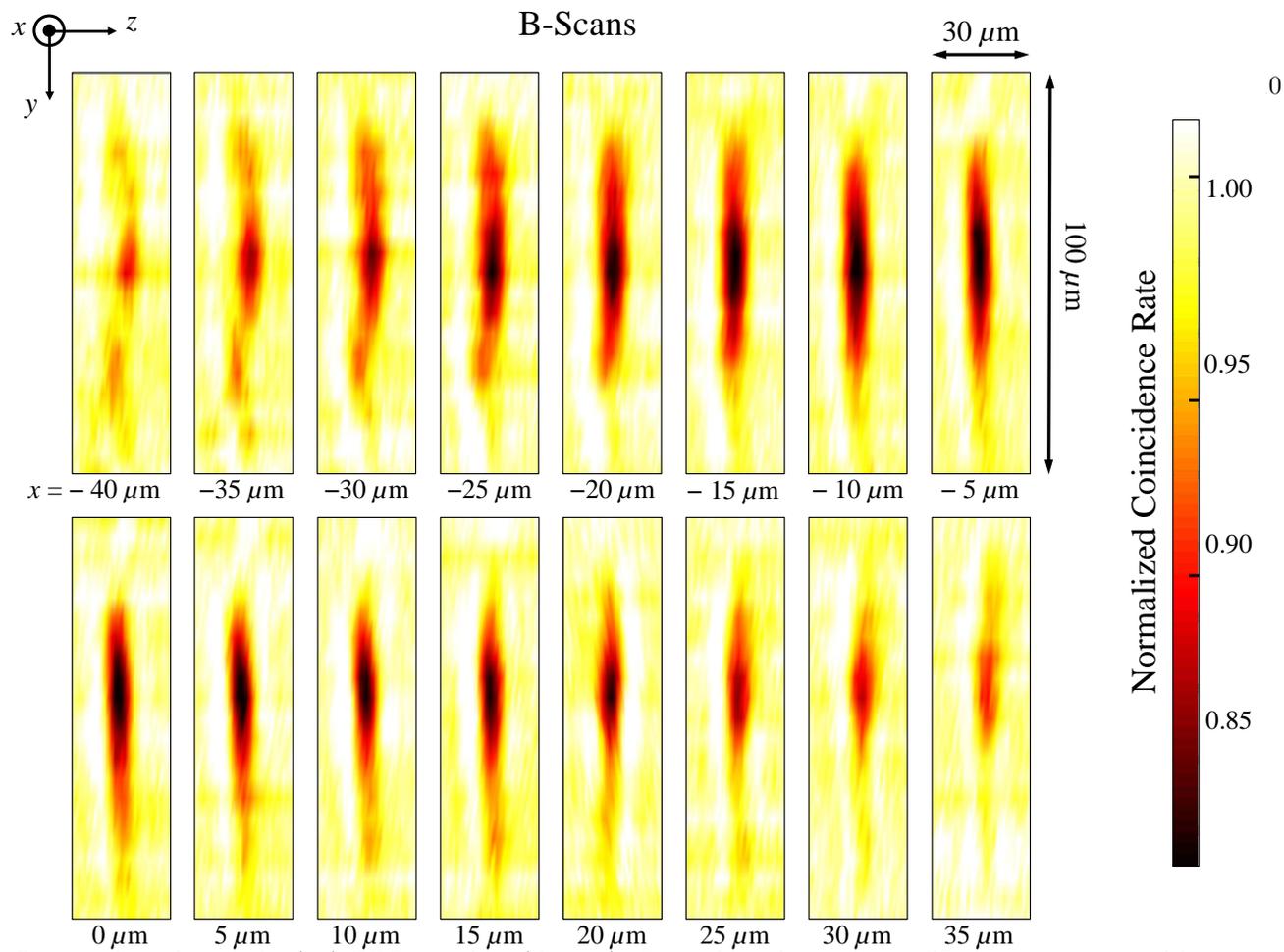}}
    \caption{Two-dimensional axial (\textit{y\,z}) QOCT sections (B-scans) of an onion-skin sample at
    different transverse positions $x$. A strong QOCT signal, representing the detection of a reflecting surface, is indicated by a decrease in the observed coincidence rate.}
    \label{QOCTofOnion-BScan(YZ)}
 \end{center}
\end{figure}

\begin{figure}[htbp]
\begin{center}
   \centerline{\includegraphics[width=9cm]{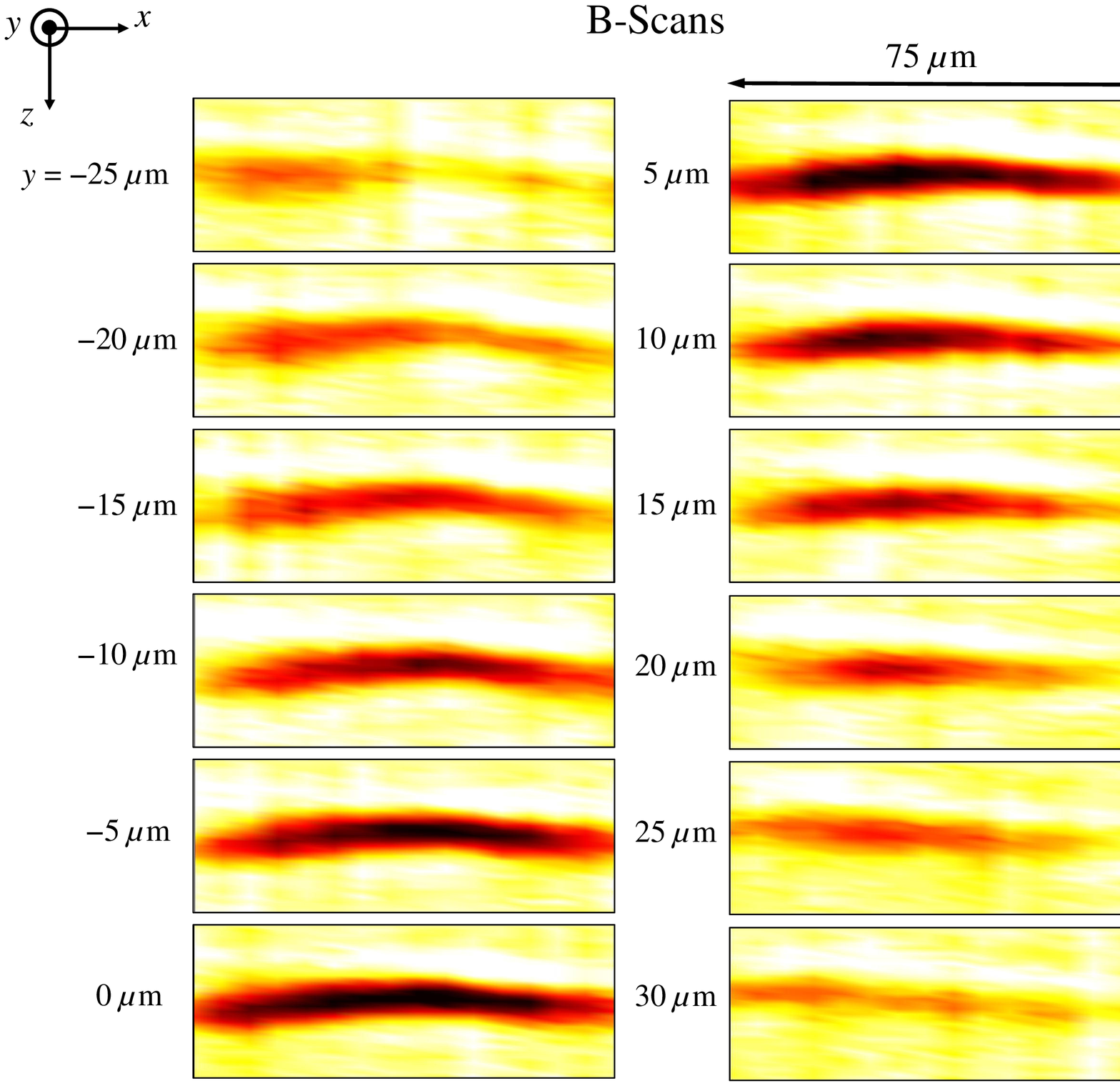}}
    \caption{Two-dimensional axial (\textit{x\,z}) QOCT sections (B-scans) of an onion-skin sample at
    different transverse positions $y$. A strong QOCT signal, representing the detection of a reflecting surface, is indicated by a decrease in the observed coincidence rate.}
    \label{QOCTofOnion-BScan(XZ)}
 \end{center}
\end{figure}

The B-scans take two forms: axial \textit{y\,z} sections at various
transverse positions $x$ and axial \textit{x\,z} sections at various
transverse positions $y$. These are displayed in
Figs.~\ref{QOCTofOnion-BScan(YZ)} and \ref{QOCTofOnion-BScan(XZ)},
respectively. The \textit{y\,z} sections are 100 $\times$ 30
$\mu\textrm{m}^2$ each, and are shown at intervals of 5
$\mu\textrm{m}$, whereas the \textit{x\,z} sections are 75 $\times$
30 $\mu\textrm{m}^2$ each, and are also shown at intervals of 5
$\mu\textrm{m}$. The \textit{x} and \textit{y} origins are placed
roughly at the center of the cell. Most of the B-scans shown in
Fig.~\ref{QOCTofOnion-BScan(XZ)} clearly display the curved
topography of the onion-skin cell surface.

\section{Discussion}

The results presented here are the first experimental QOCT data from
a biological specimen. The signal-to-noise ratio (SNR) and speed of
the QOCT technique are determined by a number of factors, including
the optical power (biphoton flux) in the
interferometer~\cite{SalehTeichSNR}. The experiments presented in
this paper were carried out with a flux of about $10^6$
biphotons/sec, corresponding to an optical power of $\approx$ 0.5
pW. This low flux, which was generated by pumping the NLC with a
pump power of about 2 mW, can be raised by simply increasing the
pump power, but an upper limit on the biphoton flux is imposed by
the saturation level of the single-photon detectors and coincidence
circuit. Faster single-photon detectors, including those relying on
the superconducting technology~\cite{SSPD(2008)}, as well as faster
coincidence circuits, are expected to greatly enhance the SNR and
speed of QOCT. Furthermore, recent advances in the production of
biphotons have provided ultrahigh axial resolution (about 1
$\mu\textrm{m}$)~\cite{Nasr(2008)}. In the offing are electrically
driven solid-state biphoton sources~\cite{Hayat(2008)} that promise
optical powers in the $\mu$W region, making QOCT even more
attractive. In short, our experiments suggest that future
enhancements in the source photon flux, spatial resolution, and
image acquisition time will help make QOCT a viable biological
imaging technique.

It is gratifying that QOCT has inspired a number of post-processing
algorithms and classical nonlinear-optical configurations that offer
dispersion-immune axial sectioning of a
sample~\cite{shapiro(2006)}--\cite{Resh(2008)}. These
``quantum-mimetic" techniques have various limitations, however, and
to date none of these proposed schemes has successfully been used to
demonstrate a scan of a biological specimen.

\section*{Acknowledgments}

This work was supported by a U.~S.~Army Research Office (ARO)
Multidisciplinary University Research Initiative (MURI) Grant; by
the Bernard M. Gordon Center for Subsurface Sensing and Imaging
Systems (CenSSIS), an NSF Engineering Research Center; and by the
Army Research Laboratory through cooperative agreement
W911NF-06-2-0040. We are grateful to A. V. Sergienko for helpful
comments, to P. G. Allen for instruction on the use of the
reflection confocal microscope, and to D. Whitney for help with the
data-collection software.

\newpage

\newpage

\newpage

\newpage

\newpage

\end{document}